\documentstyle[aps]{revtex}

\begin{document}
\author{Roberto Aquilano}
\address{Instituto de Fisica de Rosario,\\
Av. Pellegrini 250,\\
2000, ROSARIO,Argentina.}
\author{Mario Castagnino}
\address{Instituto de Astronomia y Fisica del Espacio,\\
Casilla de Correos 67, Sucursal 28,\\
1428, BUENOS AIRES, Argentina.}
\title{The Time-Asymmetry and the Entropy Gap.}
\date{January, 20th., 1996. }
\maketitle

\begin{abstract}
The universe time-asymmetry is essentially produced by its low-entropy
unstable initial state. Using quanlitative arguments, Paul Davies has
demonstrated that the universe expansion may diminish the entropy gap,
therefore explaining its low-entropy state, with respect to the maximal
possible entropy at any time. This idea is implemented in a qualitative
way\thinspace in a simple homogeneous model. Some rough coincidence with
observational data are found

$\bullet $e-mail: Castagni (a) Iafe.Edu.Ar.

$\bullet $Pacs Nrs. 05.20-y, 03.65 BZ, 05.30-d.
\end{abstract}

\section{Introduction.}

The time-asymmetry problem can be stated in the following way:

{\it How can we explain the obvious time-asymmetry of the universe and most
of its subsystems if the fundamental laws of physic are time-symmetric?}

Physicists usually answer this question first observing that, if the initial
state of the universe (or one any of its subsystems) would be an equilibrium
state, the universe (or the subsystem) will remain for ever in such state,
making impossible to find any time-asymmetry. Thus we must solve two
problems:

i.- To explain why the universe (or the subsystem)began in a non-equilibrium
(unstable, low-entropy) state, at a time that we shall call $t=0$.

ii.- To define, for the period $t>0$, a Lyapunov variable, namely a variable
that never decreases (e.g. entropy), an {\it arrow of time, }and also
irreversible evolution equations, despite the fact that the main laws of
physics are time-symmetric.

Let us comments these two problems:

i.- The set of {\it irreversible} processes that began in an unstable
non-equilibrium state constitute a {\it branch system }\cite{Reichenbach}%
{\it , }\cite{Davies}{\it . }That is to say, every one of these processes
began in a non-equilibrium state, which state was produced by a previous
process of the set. E. g.: Gibbs ink drop (initial unstable state) spreading
in a glass of water ( irreversible process) it is only possible if there was
first an ink factory which extracted the necessary energy from an engine,
where coal (initial unstable state) was burned (branched irreversible
process); in turn, coal was created with energy coming from the sun, where $%
H $ (initial unstable state) is burned (branched irreversible process);
finally, $H$ was created using energy obtained from the unstable initial
state of the universe (the absolute initial state of the branch system).
Therefore using this hierarchical chain, all the irreversible processes are
related to the cosmological initial condition, the single one that must be
explained.

ii.- Once we have understood the origin of the initial unstable state of
each irreversible process within the universe (even if we have not yet
explained the origin of the initial state of the whole universe) it is not
difficult to obtain a growing entropy (and irreversible evolution
equations), in any subsystem within the universe. With this purpose we can
consider, e. g. that forces of stochastic nature penetrate from the exterior
of each subsystem adding stochastic terms \cite{Mackey}. Alternatively,
taking into account the enormous amount of information contained in the
subsystem we can neglect some part of it \cite{Mackey},\cite{Irre}. Thirdly,
or we can use more refined mathematical tools \cite{Rigged} , \cite{Cosmo}.
With any one of this tools we can solve this problem.

It remains only one problem: Why the universe began in an unstable
low-entropy state?

If we exclude a miraculous act of creation we have only three scientific
answers:

i.- The unstable initial state of the universe is a law of nature.

ii.- This state was produced by a fluctuation.

iii.- The expansion of the universe (coupled to the nuclear reactions in it)
produces a decreasing of the (matter-radiation) entropy gap.

The first solution is only a way to bypass the problem, while the
fluctuation solution is extremely improbable. In fact, the probability of a
fluctuation diminish with the number of particles of the considered system
and the universe is the system with the largest number of particles.

The third solution was sketched by Paul Davies in reference \cite{Davies}
,only as a qualitative explanation. The expansion of the universe is like an
external agency (namely: external to the matter-radiation system of the
universe) that produces a decreasing of its entropy gap, with respect to de
maximal possible entropy, $S_{\max }$ (and therefore an unstable state), not
only at $t=0$ but in a long period of the universe evolution. We shall call
this difference the entropy gap $\Delta S,$ so the actual entropy will be $%
S_{act}=S_{\max }+\Delta S$ (fig 1.). In this essay we will try to give a
quantitative structure to Davies solution using an oversimplified
cosmological model,which, anyhow, yields a first rough numerical coincidence
with observational data.

\section{ The entropy gap.}

It is well known that the universe isotropic and homogeneous expansion is a
reversible process with constant entropy \cite{Tolman}. In this case the
matter and the radiation of the universe are in a thermic equilibrium state $%
\rho _{*}(t)$ at any time $t$. As the radiation is the only important
component, from the thermodynamical point of view, we can chose $\rho
_{*}(t) $ as a black-body radiation state \cite{COBE}, i. e. $\rho _{*}(t)$
will be a diagonal matrix with main diagonal: 
\begin{equation}
\rho _{*}(\omega )=ZT^{-3}\frac 1{e^{\frac \omega T}-1}  \label{1}
\end{equation}
where $T$ is the temperature, $\omega $ the energy, and $Z$ a normalization
constant (\cite{Landau}, eqs. (60.4) and (60.10)). The total entropy is: 
\begin{equation}
S=\frac{16}3\sigma VT^3  \label{2}
\end{equation}
(\cite{Landau}, eq. (60.13)) where $\sigma $ is the Stefan-Boltzmann
constant and $V$ a commoving volume.

Let us consider an isotropic and homogeneous model of universe with radius
(or scale) $a.$ As $V\sim a^3,$ and, from the conservation of the
energy-momentum tensor and radiation state equation, we know that $T\sim
a^{-1},$ we can verify that $S=const.$ Thus the irreversible nature of the
universe evolution is not produced by the universe expansion, even if $\rho
_{*}(t)$ has a slow time variation.

Therefore, the main process that has an irreversible nature after decoupling
time is the burning of unstable $H$ in the stars (that produces $He$ and,
after a chain of nuclear reactions, $Fe$). This nuclear reaction process has
certain mean life-time $t_{NR}=\gamma ^{-1}$ and phenomenologically we can
say the estate of the universe, at time $t$, is: 
\begin{equation}
\rho (t)=\rho _{*}(t)+\rho _1e^{-\gamma t}+0[(\gamma t)^{-1}]  \label{3}
\end{equation}
where $\rho _1$ is certain phenomenological coefficient constant in time,
since all the time variation of nuclear reactions is embodied in the
exponential law $e^{-\gamma t}$. We can foresee, also on phenomenological
grounds, that $\rho _1$ must peak strongly around $\omega _1$, the
characteristic energy of the nuclear process.

All these reasonable phenomenological facts can also be explained
theoretically: Eq. \ref{3} can be computed with the theory of paper \cite
{Sudarshan} or with rigged Hilbert space theory \cite{Rigged}. In reference 
\cite{Laura} it is explicitly proved that $\rho _1$ peaks strongly at the
energy $\omega _1$.

The normalization conditions at any time $t$ yields: 
\begin{equation}
tr\rho (t)=tr\rho _{*}(t)=1,....tr\rho _1=0  \label{4}
\end{equation}
The last equations show that $\rho _1$ is not a state but only the
coefficients of a correction around the equilibrium state $\rho _{*}(t).$ It
is explicitly proved in paper \cite{Laura}, that $\rho _1$ has a vanishing
trace.

We are now able to compute the {\it entropy gap }$\Delta S$ with respect to
the equilibrium state $\rho _{*}(t)$ at any time $t.$ It will be the
conditional entropy of the state $\rho (t)$ with respect to the equilibrium
state $\rho _{*}(t)$ \cite{Mackey}: 
\begin{equation}
\Delta S=-tr[\rho \log (\rho _{*}^{-1}\rho )]  \label{5}
\end{equation}
Using now eq. \ref{3}, and considering only times $t\gg t_{NR}=\gamma ^{-1}$
we can expand the logarithm to obtain: 
\begin{equation}
\Delta S\approx -e^{-\gamma t}tr\left( \rho _{*}^{-1}\rho _1^2\right)
\label{6}
\end{equation}
where we have used eq. \ref{4}. We can now introduce the equilibrium state 
\ref{1} for $\omega \gg T$ . Then: 
\begin{equation}
\Delta S\approx -Z^{-1}T^3e^{-\gamma t}tr(e^{\frac \omega T}\rho _1^2)
\label{7}
\end{equation}
where $e^{\frac \omega T}$ is a diagonal matrix with this function as
diagonal. But as $\rho _1$ is peaked around $\omega _1$ we arrive to a final
formula for the entropy gap: 
\begin{equation}
\Delta S\approx -CT^3e^{-\gamma t}e^{\frac{\omega _1}T}  \label{8}
\end{equation}
where $C$ is a positive constant.

\section{Evolution of the entropy gap $\Delta S$.}

We have computed of $\Delta S$ for times larger than decoupling time and
therefore, as $a\sim t^{\frac 23}$ and $T\sim a^{-1},$ we have:

\begin{equation}
T=T_0\left( \frac{t_0}t\right) ^{\frac 23}  \label{9}
\end{equation}
where $t_0$ is the age of the universe and $T_0$ the present temperature.
Then: 
\begin{equation}
\Delta S\approx -C_1e^{-\gamma t}t^{-2}e^{\frac{\omega _1}{T_0}\left( \frac{%
t_0}t\right) ^{\frac 23}}  \label{10}
\end{equation}
where $C_1$ is a positive constant. Fig. 2 is the graphic representation of
curve $\Delta S(t).$ It has a maximum at $t=t_{cr_1}$ and a minimum at $%
t=t_{cr_2}.$ Let us compute these critical times. The time derivative of the
entropy reads: 
\begin{equation}
\stackrel{\bullet }{\Delta S}\approx \left[ -\gamma -2t^{-1}+\frac 23\frac{%
\omega _1}{t_0T_0}\left( \frac{t_0}t\right) ^{\frac 13}\right] \Delta S
\label{11}
\end{equation}
This equation shows two antagonic. effects. The universe expansion effect is
embodied in the second and third terms in the square brackets an external
agency to the matter-radiation system such that, if we neglect the second
term, it tries to increase the entropy gap and, therefore, to take the
system away from equilibrium (as we will see the second term is practically
negligible). On the other hand, the nuclear reactions embodied in the $%
\gamma $-term, try to convey the matter-radiation system towards
equilibrium. These effects becomes equal at the critical times $t_{cr}$ such
that: 
\begin{equation}
\gamma t_0+2\frac{t_0}{t_{cr}}=\frac 23\frac{\omega _1}{T_0}\left( \frac{t_0%
}{t_{cr}}\right) ^{\frac 13}  \label{12}
\end{equation}
For almost any reasonable numerical values this equation has two positive
roots: $t_{cr_1}\ll t_0\ll t_{cr_2}$. Precisely:

i.- For the first root we can neglect the $\gamma t_0$-term and we obtain: 
\begin{equation}  \label{13}
t_{cr_1}\approx t_0\left( 3\frac{To}{\omega _1}\right) ^{\frac 32}
\end{equation}
(this quantity, with minus sign, gives the third unphysical root).

ii.- For the second root we can neglect the $2(t_0/t_{cr})-$term, and we
find: 
\begin{equation}
t_{cr_2}\approx t_0\left( \frac 23\frac{\omega _1}{T_0}\frac{t_{NR}}{t_0}%
\right) ^3  \label{14}
\end{equation}

\section{Numerical estimates.}

We must chose numerical values to four parameters: $\omega _1=T_{NR},$ $%
t_{NR}=\gamma ^{-1},$ $t_0,$ and $T_0.$

$T_{NR}$ and $t_{NR}$ can be chosen between the following values \cite{Cumul}%
: 
\begin{equation}
T_{NR}=10^6..to..10^8{}^0K  \label{15}
\end{equation}
\[
t_{NR}=10^6..to..10^9.years 
\]
while for $t_0$ and $T_0$ we can take: 
\begin{equation}
t_0=1.5\times 10^{10}.years  \label{16}
\end{equation}
\[
T_0=3^0K 
\]
In order to obtain a reasonable result we choose the lower bounds for $%
T_{NR} $ and $t_{NR\text{ }}$ and we obtain for $t_{cr_1}:$%
\begin{equation}
t_{cr_1}\approx 1.5\times 10^3.years  \label{17}
\end{equation}
So $t_{cr_1}$ is smaller than the decoupling time and, it must not be
considered since the physical processes before this time are different than
those we have used in our model. Also, we must consider only times $%
t>t_{NR}=\gamma ^{-1},$ in order to use eq. \ref{6}. So, only the r.h.s.
from the dashed line of fig. 2 can be taken into account.

For $t_{cr_2}$ we obtain: 
\begin{equation}  \label{18}
t_{cr_2}\preceq 10^4t_0
\end{equation}
From eqs. \ref{17} and \ref{18} we can see that $t_{cr_1}\ll t_0\ll
t_{cr_2}. $ Thus:

-From $t_{NR}$ to $t_{cr_2}$ the expansion of the universe produces a
decreasing of entropy gap, according to Paul Davies prediction. It probably
produces also a growing order, and therefore the creation of structures like
clusters, galaxies and stars \cite{Reeves}.

-After $t_{cr_2}$ we have a growing of entropy, a decreasing order and a
spreading of the structures: stars energy is spread in the universe,which
ends in a thermic equilibrium \cite{AJP}. In fact, when $t\rightarrow \infty 
$ the entropy gap vanishes (se eq. \ref{10}) and the universe reaches a
thermic equilibrium final state.

$t_{cr_2}\preceq 10^4t_0$ is the frontier between the two periods. Is the
order of magnitude of $t_{cr_2}$ a realistic one? In fact it is, since $%
10^4t_0\approx 1.5\times 10^{14}years$ after the big-bang all the stars will
exhaust their fuel \cite{AJP} , so the border between the two periods most
likely have this order of magnitude and must also be smaller than this
number. This is precisely the result of our calculations.

\section{Conclusion.}

Clearly we have not a physical argument to choose the lower bounds in eq.\ref
{15}, we have just chosen this values for convenience. Therefore, our model
is extremely naive and simplified: an homogeneous isotropic universe. In
real universe nuclear reactions take place within the stars, that only can
be properly considered in a inhomogeneous geometry. Also there are
condensation phenomena that increase $S_{\max }$, in such a way that, even
if $\Delta S$ first decreases and then increases, $S_{act}$ always
increases, according to the second low of Thermodynamics (fig .1). Notice
that we have only take into account the global expansion effect of the
universe and not the local effect of the gravitational field. Perhaps this
effect can yield better results. Nevertheless, our model is at the edge of a
correct physical prediction.

Summarizing, we have proved that Davies qualitative idea can be implemented
quantitatively, preparing the scenario for more detailed calculations.

\section{Acknowledgment.}

This work was partially supported by grants : .........................of
the European Community, PID-0150 of CONICET (National Research Council of
Argentina), EX-198 of the Buenos Aires University, and 12217/1 of
Fundaci\'on Antorchas.

\section{Figure caption.}

Fig. 1. The evolution of the maximum entropy and the actual entropy.

Fig 2. The evolution of the entropy gap. This figure is only qualitative,
the scales are not the real ones.

\end{document}